\documentclass[letterpaper]{article}

\usepackage[T1]{fontenc}

\usepackage{geometry}
\usepackage{setspace}

\usepackage[style=chem-acs, articletitle=true]{biblatex}
\usepackage{graphicx}
\usepackage{float}

\newfloat{scheme}{htbp}{los}
\floatname{scheme}{scheme}
\floatname{chart}{chart}
\newfloat{graph}{htbp}{loh}

\usepackage{chemformula} % formulas using \ch{}

\usepackage{lipsum}

\usepackage{comment}

\usepackage{hyperref}

\hypersetup{
    colorlinks=true,
    linkcolor=blue,
    filecolor=magenta,
    urlcolor=cyan,
}

\newcommand{\change}[1]{\textcolor{black}{#1}}

\usepackage{authblk}
\author[1]{Luca Leoni$^*$}
\author[1,2]{Cesare Franchini}
\affil[1]{Department of physics and astronomy "Augusto Righi", alma mater studiorum - universit\`a di Bologna, Bologna, 40127 Italy}
\affil[2]{university of Vienna, faculty of physics and center for computational materials science, Vienna, Austria}
\title{Machine-learned dynamics of surface polarons at reduced oxide surfaces}
\date{*email: luca.leoni12@unibo.it}

\begin{document}

\maketitle

\begin{abstract}
	% ~209 words, I have found other papers with similar lenght in the abstract
	Reducible oxides exhibit a rich interplay of electronic, structural, and chemical properties that underpins applications in catalysis, photovoltaics, batteries, and energy storage. This interplay is strongly shaped by excess electrons, often introduced by oxygen vacancies, that localize as small polarons and influence charge transport and surface chemistry. At surfaces, these polarons play a central role in charge localization, mobility, and reactivity, yet their finite-temperature dynamics remain difficult to access from first principles due to the long time scales needed to adequately sample polaron's hopping. To overcome this limitation, we extend machine-learning-assisted polaron dynamics to redox-active oxide surfaces, using oxygen-deficient rutile \ch{TiO2}(110) as a paradigmatic case. By accessing several nanoseconds of dynamics over a range of temperatures, we show that small-polaron mobility at the reduced rutile \ch{TiO2}(110) surface is suppressed by several orders of magnitude relative to the corresponding bulk material, providing a microscopic interpretation of the lower electron mobilities observed in porous rutile \ch{TiO2} compared with single-crystal samples. This suppressed mobility arises from the loss of favourable hopping pathways: surface polaron motion is largely confined to planar inter-row trajectories within the second topmost layers, with only rare interlayer hopping events. These results establish a transferable machine-learning strategy for investigating polaron dynamics in reducible oxides.
\end{abstract}

\section{TOC Graphic}
\rule{0.05in}{1.75in}%
\begin{minipage}[b][1.75in]{3.25in}
	\centering
	\includegraphics[height=1.75in]{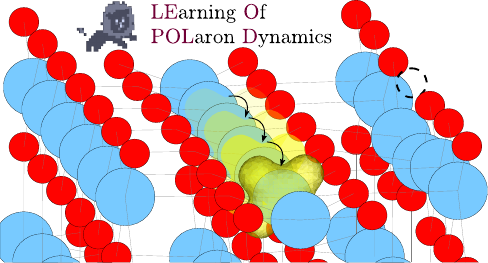}
\end{minipage}%
\rule{0.05in}{1.75in}

\newpage

%%%%%%%%%%%%%%%%%%%%%%%%%%%%%%%%%%%%%%%%%%%%%%%%%%%%%%%%%%%%%%%%%%%%%
%% Adding double columns for writing
%% NOTE: for submission this needs to be cancelled!
%%%%%%%%%%%%%%%%%%%%%%%%%%%%%%%%%%%%%%%%%%%%%%%%%%%%%%%%%%%%%%%%%%%%%
\twocolumn

%% NOTE: guidlines for letter sas to not use Headings here
%%%%%%%%%%%%%%%%%%%%%%%%%%%%%%%%%%%%%%%%%%%%%%%%%%%%%%%%%%%%%%%%%%%%%

Reducible oxide surfaces are central to a wide range of functional materials because the transition-metal cations they contain can reversibly adopt different oxidation states, enabling charge localization, defect formation, and surface redox activity that can be exploited in applications ranging from heterogeneous catalysis and photocatalysis to energy conversion and storage~\cite{Rousseau2020,GANDUGLIAPIROVANO2007219}.

The facile formation of oxygen vacancies ($V_\mathrm{O}$), enabled by their relatively low formation energies, gives rise to reduced surface centers composed of neutral oxygen vacancies and transition-metal cations with modified oxidation states. The localization of the excess electronic charge provided by $V_\mathrm{O}$ can induce the formation of small polarons, namely defect complexes in which an excess electron is trapped at a transition-metal site and strongly coupled to lattice distortions, leading to local structural modifications~\cite{Franchini2021}. In turn, these composite electron-phonon polaronic defects alter the geometric and electronic structure, strongly influencing the physical and chemical properties, and they play a decisive role in their technological applications.

Polaron formation is ubiquitous in many transition-metal oxides, including TiO$_2$~\cite{BREDOW2002417, DiValentin2005, DiValentin2008,DiValentin2011,Chiesa2013,Setvin2014,YIN201858}, CeO$_2$~\cite{GandugliaPirovano2009, Esch2025, Zhang2019, Pelatti2025}, Fe$_2$O$_3$~\cite{Iordanova2005, Carneiro2017, Pastor2019, Ahart2020, Bandaranayake2020, Redondo2024} and BiVO$_4$~\cite{Wiktor2018, Wiktor2019, Zhang2023, Liu2026}, which are among the most widely studied reduced oxide surfaces exhibiting polaronic behaviour.
In particular, titanium dioxide (\ch{TiO2}), an Earth-abundant material of steadily growing technological importance~\cite{DIEBOLD2003}, has become a key test-bed system for studying polaronic effects~\cite{Deskins2007, Setvin2014, Deak2014, Deak2015, Moses2016, Elmaslmane2018}, with applications spanning biocompatible technologies, gas sensing, photocatalysis, and photovol\-taics~\cite{Ali2018,FUJISHIMA2008,Marcelis2024}. The performance of these applications is critically governed by the nature and transport properties of polaronic charges.
After decades of study, it is now established that the two most common polymorphs of \ch{TiO2}, rutile and anatase, host small and large electron polarons, respectively~\cite{Elmaslmane2018,Dai2024}, which exhibit distinct transport regimes. Whereas large polarons behave similarly to quasi-free particles with a  metal-like temperature dependence of the conductivity, small polarons undergo thermally activated hopping transport, with mobilities that increase with increasing temperature~\cite{Zhang2007,Franchini2021}.

Experimental measurements of small polaron mobility are challenging and can depend sensitively on sample quality, reduction level, and the presence of defects such as Ti interstitials or Nb impurities.
As a result, reported values for rutile TiO$_2$ span several orders of magnitude, from $10^{-2}$ to $10$ cm$^2$V$^{-1}$s$^{-1}$~\cite{Tang1994, Yagi1996, Hendry2004}.
Importantly, it has been reported that those values drop when estimated in porous \ch{TiO2} samples by several orders of magnitude~\cite{Hendry2006}, reaching reported values as low as $7\times10^{-6}$~cm$^2$V$^{-1}$s$^{-1}$~\cite{Dittrich2000}.
This reduction has been attributed to electron immobilization at trapping sites, as well as to intrinsic mobility suppression arising from broken symmetry and the restricted transport pathways characteristic of high-surface-area porous materials.
These findings suggest that small-polaron mobility at surfaces is likely to be significantly lower than the corresponding bulk value~\cite{Hendry2006,Dittrich2000}.

First-principles simulations provide an efficient framework to interpret and elucidate the microscopic mechanisms underlying small-polaron formation and dynamics. Starting from the seminal work of Deskins et al.~\cite{Deskins2007}, several static calculations on bulk and surface rutile TiO$_2$ have been performed, providing valuable insight into the most favorable trapping sites, activation energies, selected polaron hopping directions, and the degree of adiabaticity~\cite{Janotti2013, Spreafico2014, Moses2016, Elmaslmane2018, morita2023, Shi2024}. Beyond static approaches, the actual dynamical pathways of polaron motion can be accessed through first-principles molecular dynamics (FPMD)~\cite{Kowalski2010,Setvin2014,Reticcioli2018}. However, the high computational cost of FPMD limits simulated polaron-hopping trajectories to timescales of only a few picoseconds. This unavoidably restricts the exploration of configurational space and reduces the statistical sampling needed to accurately predict mobilities.
These limitations have recently been overcome by integrating machine-learning interatomic potentials (MLIPs)~\cite{Behler2007,Bartok2010} into ab initio small-polaron dynamics with the LEOPOLD~\cite{LEOPOLD} architecture, where time-dependent polaron trajectories are learned through the dynamical prediction of the site-selective polaron occupation matrix using equivariant graph neural networks. This approach extends polaron dynamics to nanosecond timescales and enables the sampling of thousands of polaron trajectories. Applied to small-polaron hopping in bulk rutile TiO$_2$, LEOPOLD correctly reproduces the temperature-dependent hopping regime characteristic of small-polaron transport and yields a room-temperature mobility of 1.6~cm$^2$/Vs, in agreement with recent spectroscopic measurements~\cite{Hendry2006, Austin2001}.

\begin{figure}[t]
	\centering
	\includegraphics[width=\linewidth]{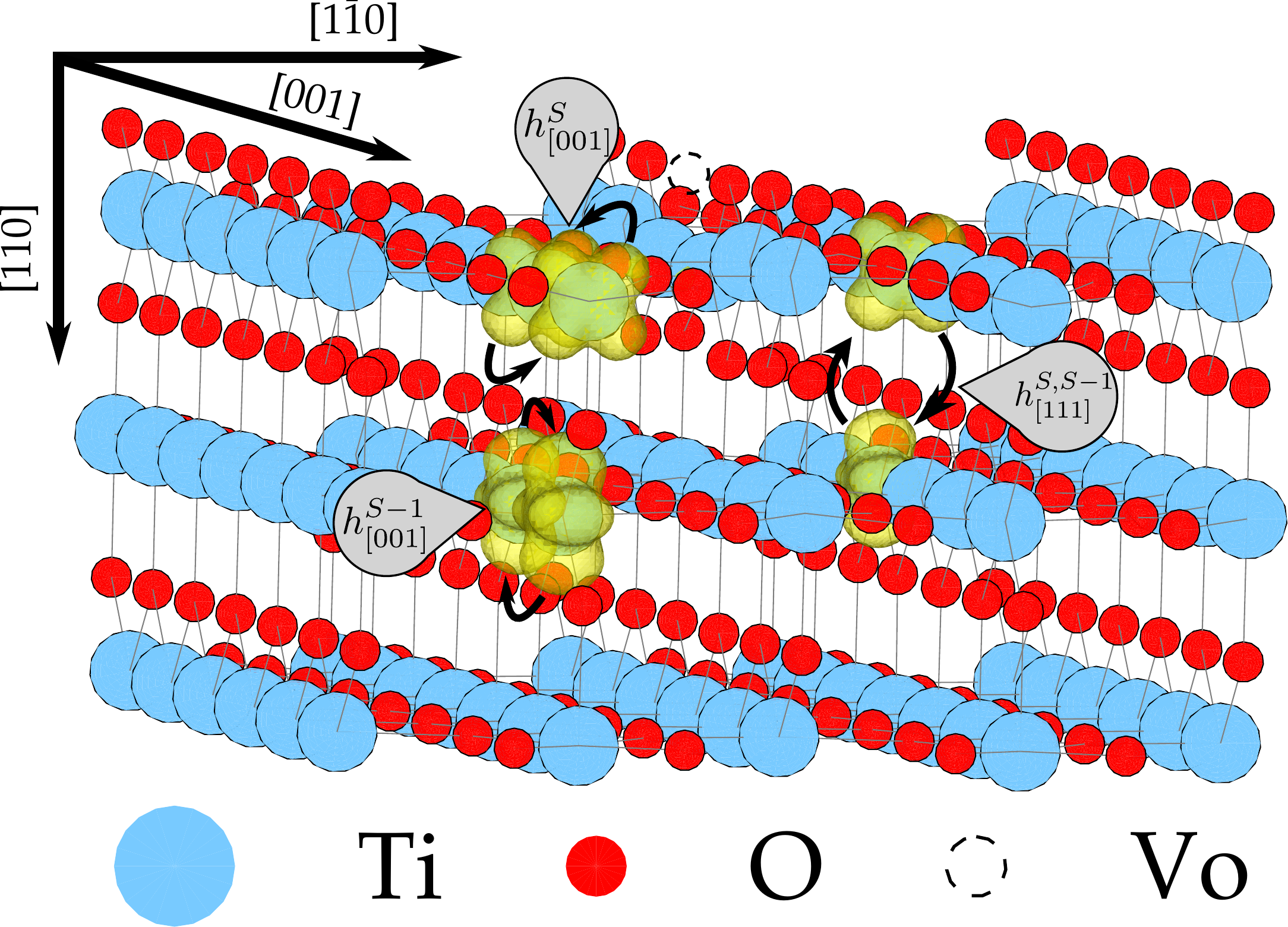}
	\caption{Representation of the \ch{TiO2} (110) surface with a surface oxygen vacancy. The most probable polaron hopping events observed during the simulation are shown, with the polaronic charge density depicted in the two adjacent states and connected by arrows. Among the hopping events reported, two occur along the preferred [001] direction: one within the surface layer $S$ (labelled $h^S_{[001]}$) and one within the subsurface layer $S-1$ (labelled $h^{S-1}_{[001]}$) directly below it. The third occurs along the [111] direction, corresponding to polaron transfer between the surface and subsurface layers (labelled $h^{S, S-1}_{[111]}$).}
	\label{fig:model}
\end{figure}

In this Letter, we extend the capabilities of LEO\-POLD to reduced oxide surfaces by predicting small-polaron mobility at the oxygen-deficient rutile TiO$_2$ (110) surface.
A comparative analysis between bulk and surface dynamics reveals \change{a reduction of four orders of magnitude} in small-polaron mobility at the surface, primarily caused by the suppression of favourable hopping pathways, which are largely confined to planar inter-row trajectories within the second topmost layers and only rarely involve interlayer hopping.
These results provide a microscopic explanation for experimental observations and broaden the applicability of ML-based ab initio simulations to the study of polaron transport in realistic oxide environments.

The structural model used in this study is shown in Fig.~\ref{fig:model}.
It consists of a 539-atom asymmetric slab containing five Ti layers in a $6\times3$ two-dimensional supercell, with one surface oxygen vacancy, $V_\mathrm{O}$, acting as an electron donor~\cite{Kowalski2010, Setvin2014, Moses2016}.
One of the two excess electrons associated with the vacancy was compensated by a uniform background charge, resulting in a single mobile small polaron.
The database for the ML-based LEOPOLD run was generated from FPMD simulations performed using the Vienna Ab initio Simulation Package (VASP)~\cite{Kresse1996, Kresse1999}, within the framework of density functional theory augmented by an effective on-site Hubbard correction of $3.9$~eV applied to the d orbital of Ti atoms~\cite{Dudarev1998, Setvin2014}.

By initializing the polaron trapping site at different locations~\cite{Reticcioli2019}, we performed several FPMD simulations at from 300 to 700~K, for a total simulation time of approximately 100~ps.
These simulations generated thousands of configurations and yielded several tens of hopping events, in which the small polaron moved from one site to a neighbouring site through a predominantly adiabatic process~\cite{Deskins2007, LEOPOLD}.
The resulting FPMD data were used as initial training dataset for the LEOPOLD model and then further enriched through an active-learning strategy used to insert more exotic jumps identified during the runs by the model itself~\cite{LEOPOLD}.
Additional computational details, including a description of the MLMD LEOPOLD architecture, are provided in the Methods section.

\begin{figure*}[t]
	\centering
	\includegraphics[width=\linewidth]{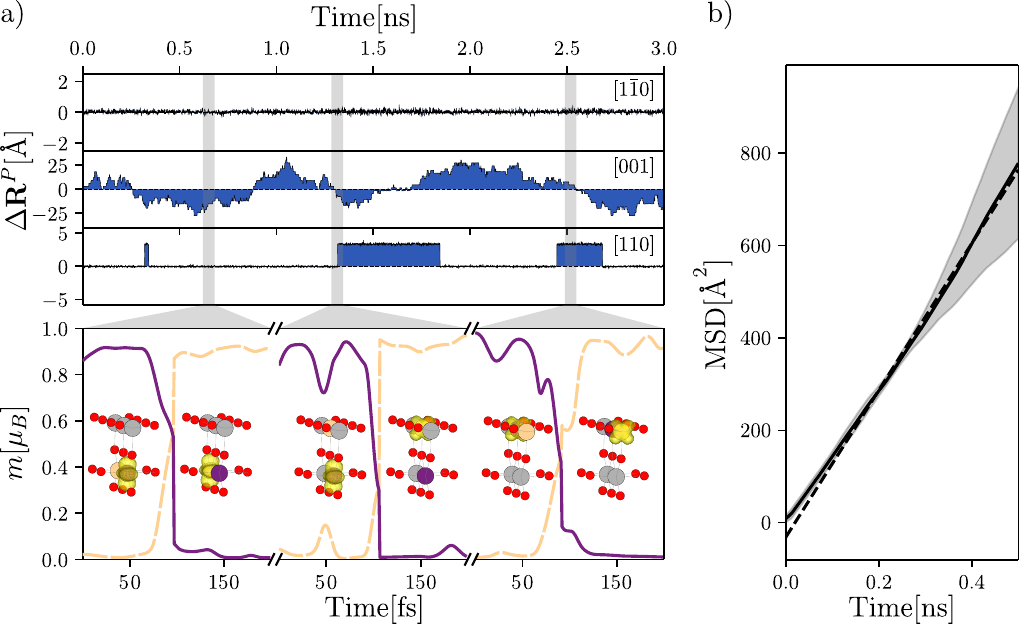}
	\caption{a) Leopold simulated small polaron dynamics on \ch{TiO2}(110) at 700~K. In the top panel the projection displacement of the polaron position, $\Delta \mathbf{R}^P$, on the three main crystallographic directions is shown as a function of the simulation time. The components show a clear discontinuous motion due to the polaron jumping between \ch{Ti} atoms mainly through jumps in the [001] direction.
		The bottom panel shows a close up on the three type of jumps present in the dynamic, showing the magnetization profiles as predicted by LEOPOLD, where the purple continuous line describes the magnetization on the donor atom and the yellow dashed line the magnetization on the receiving atom. b) The mean square displacement of the polaron averaged from three different runs at 700~K is reported. The continuous line and the dashed area represent mean MSD and deviation from the mean for the three runs, a linear fit is shown as dashed line.
	}
	\label{fig:nsrun}
\end{figure*}

A representative example of a 3~ns MLMD run obtained using the protocol described above at 700~K is shown in Fig.~\ref{fig:nsrun}(a). LEOPOLD identifies three main distinct types of hopping trajectories schematically illustrated in Fig.~\ref{fig:model} ($h^{S-1}_{[001]}$, $h^{S}_{[001]}$, and $h^{S,S-1}_{[111]}$), in agreement with previous FPMD studies~\cite{Kowalski2010, Setvin2014, Reticcioli2017, Reticcioli2018}. These trajectories are shown individually at the bottom of Fig.~\ref{fig:nsrun}(a) over shorter femtosecond time windows.
The most frequent hopping event by far, accounting for approximately 87\% of the total events, involves small-polaron transport along the [001] rows in the subsurface S-1 layer adjacent to the surface row containing $V_\mathrm{O}$, $h^{S-1}_{[001]}$. In about 12\% of the cases, the polaron hops along the surface Ti row next to $V_\mathrm{O}$, $h^{S}_{[001]}$. Interlayer hopping events are observed only very rarely, approximately 1\% of the time, and involve electron transfer from S-1 to S, or vice versa, $h^{S,S-1}_{[111]}$. Small-polaron transport to deeper layers was never observed, confirming the clear tendency of small polarons to form and remain confined within the topmost surface layers~\cite{Papageorgiou2010, Setvin2014, Reticcioli2019,Reticcioli2022}. Inter-row hopping between adjacent Ti rows, which are $\sim$3~\AA~ apart and connected by oxygen atoms is equally rare.

Each individual hopping event is tracked by monitoring the time evolution of the local spin magnetization, $m$~\cite{Janotti2013,LEOPOLD}, as shown in the bottom panels of Fig.~\ref{fig:nsrun}(a). Initially, the polaron is fully localized on a lattice site, with a projected local spin moment of $\sim 0.9~\mu_B$ (solid violet curve). During the hopping process, the spin moment on the initial site progressively decreases, while the increasing probability of finding the polaron on the neighboring final site leads to a corresponding increase of the local spin moment on that site (dashed orange curve). At the end of the hopping event, the small polaron is fully localized on the final site, which exhibits the local spin moment of $\sim 0.9~\mu_B$.
The point at which the two curves intersect is generally identified as the transition state~\cite{Deskins2007}.

The observed hopping trajectories are associated with a sudden jump in the polaron displacement $\Delta\mathbf{R}^P$, which measures the spatial distance of the polaron from its starting position.
The resulting time-dependent evolution of $\Delta\mathbf{R}^P$ is collected in the top part of Fig.~\ref{fig:nsrun}(a) projected over the three main surface directions.
The statistically representative data at 700~K clearly show that polaron transport occurs preferentially along the [001] direction.
The polaron moves back and forth along the \(S-1\) Ti row with occasional hops from \(S-1\) to \(S\) and vice-versa, as evidenced by a 2.5~\AA\, displacement along the [110] direction.
The occurrence of only six \(h^{S,S-1}_{[111]}\)-type jump over a 3~ns trajectory indicates that such events are much rarer than the dominant \(h^{S}_{[001]}\) and \(h^{S-1}_{[001]}\) jumps. This large difference in probability can be rationalized from the bottom panel of Fig.~\ref{fig:nsrun}(a): the \([111]\) jump involves not only a longer distance between the initial and final Ti sites, but also a reorientation of the polaron orbital. Both factors reduce the overlap between the initial and final states, thereby strongly suppressing the hopping rate, as discussed in previous works~\cite{Deskins2007,LEOPOLD}. As a result, this type of jump remains a rare event even at high temperature.

To collect sufficient statistics for an accurate estimate of the polaron mobility, \(\mu\), four independent simulations, initialized with different polaron sites, were performed at each target temperature (300, 400, 500, 600, and 700~K). The simulation time for each run varied from 1--3~ns at 700~K, where hopping events were frequent and the model was less stable, to 9~ns at 300~K, where the polaron remained essentially localized.
For every run we obtained an estimate of the polaron diffusion coefficient, $D$, by fitting the long time regime of the polaron mean square displacement (MSD) (see Fig.~\ref{fig:nsrun}(b)), and then used the Einstein relation to evaluate $\mu$.
The final value of $\mu$ at a target temperature \change{was} then taken as the average of the mobility of the different runs.
\begin{figure}[t]
	\centering
	\includegraphics[width=\linewidth]{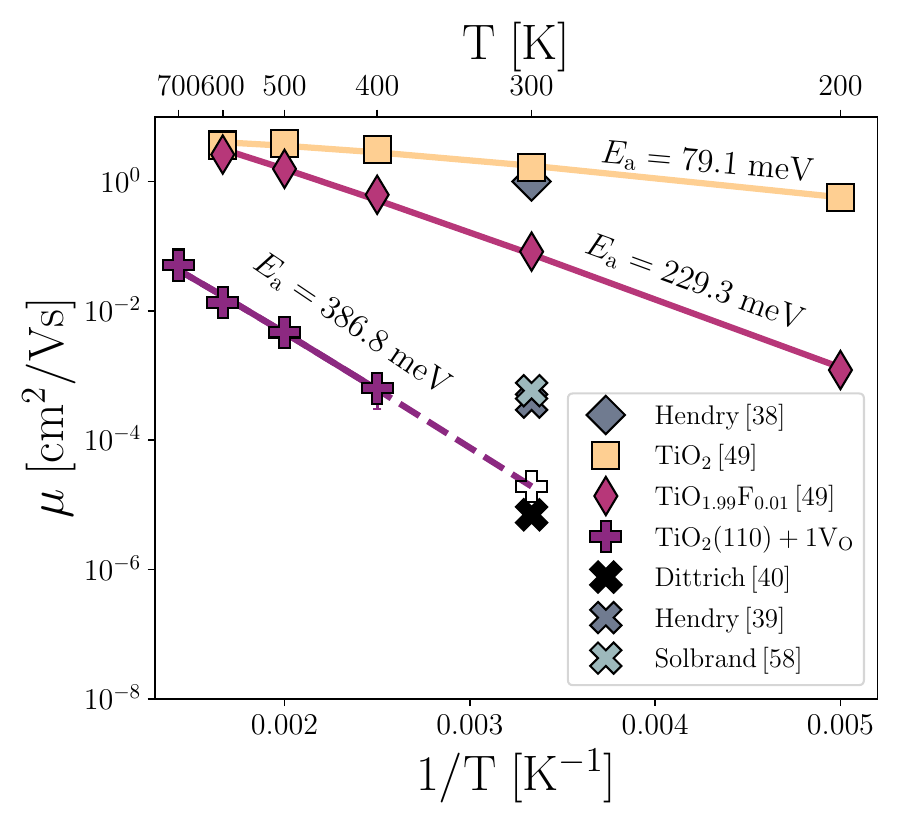}
	\caption{Polaron mobility on the oxygen-deficient rutile-\ch{TiO2}(110) surface. For comparison, the corresponding polaron mobility values in bulk rutile-\ch{TiO2}, taken from Ref.~\cite{LEOPOLD}, are also shown. The dashed purple line indicates the extrapolated region of the fit used to estimate the mobility at 300~K, shown as an open cross. Filled crosses denote different experimental measurements on porous \ch{TiO2}~\cite{Dittrich2000, Hendry2006, Solbrand1997}, while the filled rotated square a bulk experimental mobility~\cite{Hendry2004}.}
	\label{fig:mobility}
\end{figure}

In Fig.~\ref{fig:mobility}, we report the polaron mobility obtained for the reduced rutile-\ch{TiO2} surface, together with previous LEOPOLD estimates of \(\mu\) for pristine bulk \ch{TiO2} and for F-doped bulk \ch{TiO2}~\cite{LEOPOLD}. The results show that polaron mobility is strongly suppressed at the surface, decreasing by several orders of magnitude at all temperatures. This reduction is consistent with the difference between the room-temperature electron mobilities measured experimentally in single-crystal \ch{TiO2}~\cite{Hendry2004} and porous \ch{TiO2}~\cite{Hendry2006,Dittrich2000}.
We note that a direct quantitative estimate of \(\mu\) at \(300~\mathrm{K}\) was not possible within LEOPOLD, owing to the extremely low polaron mobility: no hopping event was observed over a total simulation time of \(\sim 40~\mathrm{ns}\).
Therefore, to compare our model with experiment, \change{we extrapolate the mobility to room temperature assuming Arrhenius behaviour, obtaining $\mu(300~\mathrm{K})=(1.9\pm0.8)\times10^{-5}~\mathrm{cm^2,V^{-1},s^{-1}}$. The quoted uncertainty is obtained by propagating the statistical uncertainties of the fitted activation energy and prefactor. It does not account for possible systematic deviations from Arrhenius behaviour outside the simulated temperature range. In particular, at lower temperatures, quantum delocalization, tunnelling, or a crossover toward more coherent polaron transport may become relevant~\cite{Mishchenko2015,Luo2025}; these effects are not captured by the present localized-polaron molecular-dynamics framework and could alter the extrapolated mobility. The resulting value lies within, although toward the lower end of,} the experimental range of \(7 \times 10^{-6}\)--\(5 \times 10^{-4}~\mathrm{cm^2\,V^{-1}\,s^{-1}}\) reported for electron mobilities in porous \ch{TiO2}~\cite{Dittrich2000,Solbrand1997,Hendry2006}. Furthermore, the activation barrier of \(386{\change{\pm 22}}~\mathrm{meV}\) obtained from the Arrhenius fit of the LEOPOLD mobilities is comparable to recent infrared-spectroscopy estimates of \(300\)--\(330~\mathrm{meV}\) for polaronic states in rutile \ch{TiO2}(110)~\cite{Idriss2025}.
The residual differences between our calculated mobilities and experimental values are likely related to factors not explicitly included in the present model, such as variations in vacancy concentration, disorder, morphology, and multi-polaron interactions.
In particular, the \change{strong carrier-concentra\-tion dependence of the electronic diffusion constant observed experimentally~\cite{Kopidakis2000, Yagi1996, Tang1994}, together with the established role of polaron--polaron interactions in influencing charge localization~\cite{Birschitzky2022, Birschitzky2024, Reticcioli2018}, indicates that the presence of multiple polarons may substantially alter site occupations, preferred hopping pathways, and the effective vacancy--polaron interaction, thereby affecting the calculated mobility. The mobility obtained in the present single-polaron calculations should therefore be regarded as representative of the dilute-polaron limit. Extending the present approach to interacting multipolaron systems and systematically investigating the effects of increasing oxygen-vacancy and carrier concentrations represent important directions for future work.}
% the experimentally observed dependence of the electronic diffusion constant on the carrier concentration~\cite{Kopidakis2000} points to the treatment of interacting polarons as an important direction for future developments.

\begin{figure}[t]
	\centering
	\includegraphics[width=\linewidth]{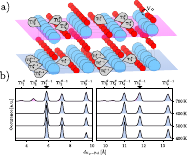}
	\caption{a) Structural model of the reduced rutile-\ch{TiO2}(110) surface, with selected polaron-trapping Ti sites labelled according to their position relative to the oxygen vacancy, \(\mathrm{V_O}\).
	b) Distributions of the polaron--vacancy distance, \(d_{\mathrm{V_O}\text{-}\mathrm{Pol}}\), obtained from independent LEOPOLD dynamics at different temperatures and initialized from different polaron sites, as indicated by the pointers in the top of the \(x\)-axis.
	The left panel corresponds to trajectories initialized with the polaron on site $\mathrm{Ti}_1^{S-1}$, whereas the right panel corresponds to trajectories initialized on $\mathrm{Ti}_5^{S-1}$ with the polaron farther from \(\mathrm{V_O}\).
	The light-blue shaded region denotes contributions from polaron localization in the \(S-1\) layer, while the pink shaded region denotes contributions from localization in the \(S\) layer.
	}
	\label{fig:vopdist}
\end{figure}

A further analysis of the generated polaron trajectories was performed to investigate the role of the interaction between the negatively charged polaron and the positively charged oxygen vacancy at the surface considering the different Ti sites shown in Fig.~\ref{fig:vopdist}(a). In particular, additional MD simulations were carried out by initializing the polaron either in a Ti row close to the vacancy or in a more distant row, and subsequently monitoring the distribution of the polaron--vacancy distance, \(d_{\mathrm{V_O}\text{-}\mathrm{Pol}}\), at the target temperatures.

The results shown in Fig.~\ref{fig:vopdist}(b) clearly indicate that the polaron is attracted to the oxygen vacancy, in agreement with previous first-principles and neural-network studies~\cite{Reticcioli2018,Birschitzky2022,Birschitzky2024}. However, the strength and character of this attraction exhibit qualitative changes with temperature and depend on the initial polaron position, as discussed below.

At low temperature (400~K), the distribution of \(d_{\mathrm{V_O}\text{-}\mathrm{Pol}}\) exhibits a pronounced decay in peak intensity with increasing polaron--vacancy separation. This indicates that the polaron spends most of the simulation time on Ti sites close to the oxygen vacancy, while thermal activation is insufficient to promote hopping toward more distant rows. In this regime, the vacancy therefore acts as an effective trapping center, and the dynamics are dominated by local exploration of the nearby Ti sublattice rather than by long-range diffusion.

This localization effect becomes much less pronounced when the polaron is initialized on a Ti row farther from the vacancy, as shown in the right panel. The reduced bias toward short \(d_{\mathrm{V_O}\text{-}\mathrm{Pol}}\) values demonstrates that LEOPOLD captures not only the attractive character of the polaron--vacancy interaction, but also its rapid spatial decay. In other words, the vacancy strongly affects the local hopping landscape, but its influence does not extend uniformly across the surface slab. Increasing the temperature progressively weakens this trapping effect, as evidenced by the enhanced occupation of configurations with larger polaron--vacancy separations. This reflects the competition between the energetic preference for vacancy-proximal sites and the entropic gain associated with accessing a larger number of distant configurations.

The time spent by the polaron in the outermost surface layer remains a small fraction of the total trajectory, as indicated by the pink shaded regions, and becomes appreciable only at \(700~\mathrm{K}\) for both initializations. Nevertheless, these rare surface-localized events provide useful insight into the near-surface trapping landscape. For trajectories initialized in the row adjacent to the vacancy, the surface contribution is concentrated mainly on the Ti\(_1^S\) and Ti\(_2^S\) sites, with Ti\(_2^S\) being significantly more populated. This indicates that the attractive interaction with the vacancy persists at the surface, but that the nearest surface Ti site, Ti\(_1^S\), is not necessarily the most favorable trapping site. The reduced occupation of Ti\(_1^S\) suggests that short-range structural relaxation, orbital orientation, and local coordination effects can overcome the purely electrostatic tendency to localize as close as possible to the positively charged vacancy. This behavior is consistent with previous first-principles studies of polaron dynamics on rutile-\ch{TiO2}(110)~\cite{Reticcioli2018}.

Finally, the fraction of time spent in the surface layer depends strongly on the row in which the polaron is initially localized: it reaches about \(17\%\) when the polaron is initialized in the row closest to the vacancy, but only about \(7\%\) when it starts from the more distant row. This difference indicates that the \change{electrostatic attraction between the negatively charged polaron and the positively charged} vacancy
% \sout{not only attracts the former laterally, but also facilitates occa\-sional hops toward the surface.} 
\change{not only draws the polaron laterally toward the defect but also lowers the local free-energy cost of occasional hops toward the surface. Thermal fluctuations provide the activation for these events, whereas the vacancy reshapes the underlying energy landscape and enhances their probability. The effect decreases rapidly with increasing distance from the vacancy, consistent with the electrostatic character of the interaction and with previous studies~\cite{Birschitzky2024,Reticcioli2018}.}
The oxygen vacancy therefore reshapes the polaron free-energy landscape in two coupled ways: it stabilizes vacancy-proximal trapping sites and enhances the probability of surface localization, while increasing temperature partially restores configurational sampling over more distant subsurface sites.
\change{ Previous studies \cite{Reticcioli2018} have also shown that repulsive polaron--polaron interactions can produce effects similar to those associated with polaron--vacancy interactions, leading to longer polaron residence times at the surface as the oxygen-vacancy concentration increases. This finding suggests that explicitly accounting for multipolaron dynamics is essential for quantitatively describing \ch{TiO2} surfaces at higher degrees of reduction and therefore represents an important direction for future work. }

In conclusion, we have extended the use of machine learning polaron dynamics to surfaces, using the LEOPOLD machine-learning framework to access the long-time dynamics of small polarons at reduced oxide surfaces, focusing on oxygen-deficient rutile-\ch{TiO2}(110) as a paradigmatic case.
The model was trained on an initial set of first-principles molecular-dynamics trajectories, covering several temperatures and subsequently enriched through active learning based on machine-learning-generated configurations.
This strategy enabled the generation of statistically meaningful polaron trajectories over nanosecond time scales, allowing us to estimate the surface polaron mobility over the temperature range 300--700~K.
Our results show that polaron transport at the reduced rutile-\ch{TiO2}(110) surface is strongly suppressed compared with the corresponding bulk case, with mobilities reduced by several orders of magnitude across the investigated temperature range.
This reduction provides a microscopic interpretation of the experimentally observed difference between electron mobilities in single-crystal and porous rutile \ch{TiO2}.

By extrapolating the Arrhenius behavior to room temperature, we obtained a mobility in quantitative agreement with the experimental range reported for porous \ch{TiO2}.
Moreover, the activation barrier extracted from the LEOPOLD trajectories is within approximately \(15\%\) of recent infrared-spectroscopy estimates for polaronic states at rutile-\ch{TiO2}(110), supporting the reliability of the present approach.

Beyond the quantitative mobility estimates, our simulations clarify the microscopic role of the oxygen vacancy in shaping surface polaron dynamics.
The vacancy acts as an attractive center for the excess electron, biasing the polaron distribution toward nearby Ti sites and promoting occasional hoppings from the most favorable subsurface trapping layer to the outermost surface layer.
This behavior indicates that surface transport is controlled not only by the local hopping barriers, but also by the defect-induced reshaping of the polaron free-energy landscape.

Overall, this work provides the first nanosecond-scale microscopic account of small-polaron mobility at a reduced oxide surface. More broadly, it shows that machine-learning architectures capable of tracking coupled lattice and charge dynamics can extend first-principles accuracy to complex realistic environments and experimentally relevant time scales.
The approach introduced here is readily transferable to other reducible oxide surfaces, defect configurations, and adsorbate environments, opening the way to predictive simulations of excess-charge dynamics in (pho\-to)catalytic, and energy-conversion materials.

\section{Methods} \label{sec:methods}

\textit{First-principles.} All first-principles calculations were performed with the GPU implementation of VASP 6.5.1~\cite{Kresse1996,Kresse1999}, including spin polarization and using the Perdew--Burke--Ernzerhof exchange-correlation functional~\cite{Perdew1996}. Electronic correlation on Ti \(3d\) states was treated within the rotationally invariant DFT+\(U\) formalism of Dudarev~\cite{Dudarev1998}, employing an effective on-site Hubbard parameter of \(3.9~\mathrm{eV}\), consistent with previous studies~\cite{Setvin2014}.

The reduced rutile-\ch{TiO2}(110) surface was modeled using a five-layer slab separated by \(15~\text{\AA}\) of vacuum and a \(6\times3\) surface supercell.
An asymmetric slab geometry was adopted, with the three topmost layers allowed to relax during the simulations, while the remaining layers were kept fixed.
A surface oxygen vacancy was introduced to generate the excess charge necessary for enabling  polaron formation.
To model a single small polaron, one of the two excess electrons released by the vacancy was compensated by a homogeneous positive background charge.
The resulting system contained 539 atoms, with surface lattice parameters of
\(19.9~\text{\AA}\) and \(18.2~\text{\AA}\)  and a cell height of \(31~\text{\AA}\).
Such model is shown in Fig.~\ref{fig:model}, where only the first three surface layers and the two rows closest to the vacancy are shown for convenience.

Standard PAW potentials were used for Ti, together with a soft O potential, allowing a plane-wave kinetic-energy cutoff of \(400~\mathrm{eV}\). Brillouin-zone sampling was restricted to the \(\Gamma\) point. Both first-principles molecular dynamics (FPMD) and machine-learning molecular dynamics (MLMD) simulations were carried out in the canonical \(NVT\) ensemble using a Nosé--Hoover thermostat~\cite{Nose1984} and a time step of \(1~\mathrm{fs}\).

\begin{figure}[t]
	\centering
	\includegraphics[width=0.99\linewidth]{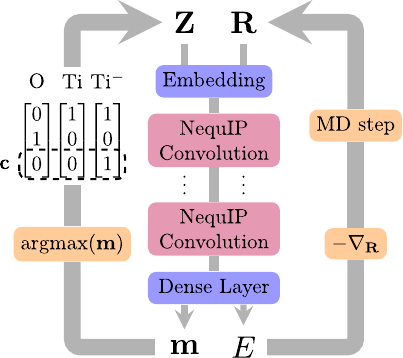}
	\caption{Graphical representation of the LEOPOLD loop. The model takes as input the polaron encoded species of the system $\mathbf{Z}$ and their atomic positions $\mathbf{R}$ to predict the polaron-aware energy and atomic magnetizations. Afterwards, the forces are computed through numerical gradient of energy and used to perform the usual molecular dynamics step to update $\mathbf{R}$.
		To evolve in time the polaronic charge instead, LEOPOLD uses the predicted $\mathbf{m}$ to find the new polaronic site as the one with the current maximum magnetization, thus allowing for a dynamic modification of $\mathbf{c}$.
	}
	\label{fig:leoarch}
\end{figure}

\textit{LEOPOLD.} The LEOPOLD package~\cite{LEOPOLD} was used both to train the polaron-aware machine-learning potential and to perform the corresponding MLMD simulations. The architecture of the model is schematically shown in Fig.~\ref{fig:leoarch}. It is based on a custom implementation of the Neural Equivariant Interatomic Potential (NequIP) framework~\cite{Batzner2022}, extended to explicitly account for the presence of a localized polaron through a dedicated ``polaron encoding''.

In conventional machine-learning interatomic potentials, atoms are typically distinguished only by their chemical species, for instance through a one-hot encoding of the nuclear charge \(Z_i\). Such a representation does not distinguish between different oxidation states of the same element. LEOPOLD overcomes this limitation by augmenting the atomic representation with an additional one-hot charge-state vector, \(\mathbf{c}\), which encodes the discrete change in local oxidation state associated with polaron localization.
In this way, the model can distinguish, for example, a regular Ti site from a Ti site hosting the excess charge, corresponding to a formal oxidation-state change of \(\pm 1\).
To assign the charge state of each configuration, we use the per-atom spin moment \(m_i\) as a descriptor of the polaron population on atom \(i\). The polaron is assigned to the atom carrying the largest absolute magnetization according to
\begin{equation}
	\label{eq:chgstate}
	c_i =
	\begin{cases}
		1 & i=\operatorname*{arg\,max}_j |m_j|, \\
		0 & \mathrm{otherwise}.
	\end{cases}
\end{equation}
This procedure yields a discrete vector \(\mathbf{c}\) that encodes the oxidation-state information of the atoms and, consequently, the instantaneous polaron position.

The resulting ``polaron encoding'' is constructed by appending \(\mathbf{c}\) to the atomic-species vector \(\mathbf{Z}\), thereby increasing the dimensionality of the input representation. For the \ch{TiO2} system considered here, this leads to three effective atomic environments: oxygen, regular titanium, and polaronic titanium. These are encoded as
\begin{equation}
	\mathrm{O}=[0,1,0], \qquad
	\mathrm{Ti}=[1,0,0], \qquad
	\mathrm{Ti}^{\mathrm{pol}}=[1,0,1].
\end{equation}
This representation enables the potential to predict energies and forces in the presence of localized excess charges, while also describing their motion.

To allow the charge-state vector \(\mathbf{c}\) to evolve dynamically during MLMD simulations, LEOPOLD also learns the atomic charge occupation, defined from the trace of the on-site occupation matrix,
\begin{equation}
	n_{i}^{\sigma l}
	=
	\sum_{jm}
	f_j
	\left\langle
	\psi_j^\sigma
	\middle|
	\mathrm{P}^{i}_{lmm}
	\middle|
	\psi_j^\sigma
	\right\rangle,
\end{equation}
where \(\psi_j^\sigma\) are the Kohn--Sham states, \(f_j\) are their occupations, \(\sigma\) denotes the spin channel, and \(\mathrm{P}^{i}_{lmm'}\) is the projector onto the \(lm\) and \(lm'\) atomic orbitals centered on atom \(i\).
In the present work, we use projectors onto the Ti \(d\) states, where the polaron is localized, and therefore set \(l=2\) throughout. For compactness, we denote the corresponding occupations as \(n_i^\sigma\).
From these spin-resolved occupations, one can obtain the projected atomic charge and magnetization as \(n_i^\uparrow+n_i^\downarrow\) and \(n_i^\uparrow-n_i^\downarrow\), respectively.

In the current implementation of LEOPOLD, the inference of the magnetization has been simplified with respect to the original formulation by training the model to predict directly the sum and difference of the spin occupations. In this way, the network effectively learns the projected on-site charges and magnetizations. This allows the model to infer the polaron population for a given atomic configuration and subsequently apply Eq.~(\ref{eq:chgstate}) to determine the charge-state vector \(\mathbf{c}\), thereby predicting the instantaneous charge state of the system.

As illustrated in Fig.~\ref{fig:leoarch}, this procedure can be embedded directly into a molecular-dynamics loop. At each time step, the model updates the charge-state vector \(\mathbf{c}\) from the instantaneous atomic configuration, using the polaron position from the previous step as an additional input. The resulting vector \(\mathbf{c}\) is a one-hot representation of the polaron position and is propagated self-consistently with the atomic degrees of freedom, allowing the polaron to migrate between atomic sites during the dynamics.

The charge state is assigned from \(\mathrm{argmax}(\mathbf{m})\), where \(\mathbf{m}\) is the predicted site-resolved magnetization. This discrete projection maps the continuous magnetization output onto a one-hot charge-state vector, ensuring that exactly one Ti site is identified as polaronic at every time step. Consequently, the dynamics remain constrained to the single-polaron manifold, even when the sum of the predicted magnetizations is not exactly unity, thereby enforcing conservation of the polaron charge throughout the simulation.

Since the LEOPOLD package is fully implemented in Python using the JAX library~\cite{jax2018}, the molecular-dynamics loop was also implemented in JAX, using JAX-MD~\cite{jaxmd2020}. This provides a fully differentiable and accelerator-compatible workflow in which the atomic structure and the polaronic charge state are updated on the fly during the simulation.

In addition, the latest version of LEOPOLD includes a custom implementation of the \(NVT\) integration scheme in JAX-MD, with support for selective dynamics. This functionality was specifically introduced to enable simulations of slab geometries such as the reduced rutile-\ch{TiO2}(110) surface considered here.

\textit{Database construction.}
The database was constructed following the strategy introduced in Ref.~\cite{LEOPOLD}. Four initial configurations were generated by imposing ad hoc local distortions around selected Ti sites, followed by structural relaxation, in order to localize the polaron in different rows and layers of the slab. Each configuration was then used as the starting point for first-principles molecular-dynamics simulations at \(300\), \(500\), and \(700~\mathrm{K}\). After thermalization through a temperature ramp of \(0.3~\mathrm{K/fs}\), each trajectory was propagated for at least \(8~\mathrm{ps}\).

In total, approximately \(100~\mathrm{ps}\) of FPMD trajectories were collected, containing 50 polaron-hopping events across the sampled temperatures. These data were used to construct the initial database, split into training and validation sets of 4963 and 1256 configurations, respectively. In both sets, \(30\%\) of the configurations were selected in the vicinity of a hopping event, ensuring that the model was exposed to the high-energy transition regions relevant for polaron transport.

The LEOPOLD potential trained on this initial dataset was then employed to perform MLMD simulations of polaron dynamics.
Configurations generated along MLMD trajectories, particularly those associated with additional hopping events, were subsequently used to augment the training database in an active-learning procedure.
The final production model contained approximately \(4.1\times10^4\) trainable parameters and was trained on the augmented dataset, comprising 6898 training and 1793 validation configurations.
The resulting training errors for forces and magnetizations were 20~meV/\AA~and 3.5~m\(\mu\), while the final validation errors were 24~meV/\AA~and 3.6~m\(\mu\).
The training specifics are reported in the supporting information.

%%%%%%%%%%%%%%%%%%%%%%%%%%%%%%%%%%%%%%%%%%%%%%%%%%%%%%%%%%%%%%%%%%%%%
%% Supporting information
%% A listing of the contents of each file supplied as Supporting Information
%% should be included. For instructions on what should be included in the
%% Supporting Information as well as how to prepare this material for
%% publications, refer to the journal's Instructions for Authors.
%%%%%%%%%%%%%%%%%%%%%%%%%%%%%%%%%%%%%%%%%%%%%%%%%%%%%%%%%%%%%%%%%%%%%
\section*{Supporting information}

\begin{itemize}
	\item supp.pdf: Accurate description of the machine learning training with specifics for the hyperparameters used. Explicit report of the final model errors and computational performances.
\end{itemize}

\section*{Data availability statement}

The new version of LEOPOLD used in this study is openly available in https://github.com/QuantumMate\-rialsModelling/leopold.
The training data underlying this study are openly available at https://doi.org/10.5281/ze\-nodo.20647088.

\section*{Acknowledgements}

The authors acknowledge support by the  National Recovery and Resilience Plan (NRRP), Mission 4 Component 2 Investment 1.3 - Project NEST (Network 4 Energy Sustainable Transition) and CN-HPC grant no. (CUP) J33C22001170001, SPOKE 7, of Ministero dell’Università e della Ricerca (MUR), funded by the European Union – NextGenerationEU. This research was partially funded by the Austrian Science Fund (FWF) 10.55776/F8100 project TACO. We acknowledge the CINECA award under the ISCRA initiative, for the availability of high-performance computing resources and support.

%%%%%%%%%%%%%%%%%%%%%%%%%%%%%%%%%%%%%%%%%%%%%%%%%%%%%%%%%%%%%%%%%%%%%
%% BIBLIOGRAPHY
%%%%%%%%%%%%%%%%%%%%%%%%%%%%%%%%%%%%%%%%%%%%%%%%%%%%%%%%%%%%%%%%%%%%%
\printbibliography

\end{document}

\maketitle

\section{Machine learning training specifics}

The training of the polaron-aware machine learning model was performed using the LEOPOLD software version 1.0.0 which has been modified respect to the original version in order to straightforward its use.
The original version of the model~\cite{LEOPOLD} was constructed as a standard NequIP architecture~\cite{Batzner2022} with an additional output layer used to predict the trace of spin up and down on-site occupation matrix $n^{\sigma}_i$, defined as in Eq.~3 of the main text.
Subsequently $n^{\sigma}_i$ is used to estimate the orbital projected magnetization at site $i$ as $m_i = n^\uparrow_i - n^\downarrow_i$.
In the new version of LEOPOLD such extra step has been eliminated by using the extra dense layer to directly predict the orbital projected on-site magnetization $m_i$ and charge $q_i$.
Such decision makes the construction of the dataset more accessible since it no longer requires to store the trace of on-site occupation matrix to train the model needing instead the angular projected magnetization and charges, quantities accessible through any standard DFT code.

Another change brought in the new version of LEOPOLD is the use of configuration files in order to specify model, dataset, and training hyper-parameters, allowing for a better reproducibility of the results.
Therefore, we report in the following the yaml configuration file used to train all the LEOPOLD models used during this study:
\begin{lstlisting}[style=yaml]
datasets:
  data_paths:
    train: Train.xyz
    validation: Valid.xyz
    test: Test.xyz
  labels:
    scalar:
      energy: energy
    vector:
      charges: charge
      forces: forces
      magmoms: magmoms
general:
  name: LEOPOLD-DATASET-TDEFAULT-H16
  device: gpu
  models_dir: checkpoints
  result_dir: results
  seed: 42
  use_float64: true
model:
  r_cutoff: 5.0
  hidden_irr: 42x0e + 8x1e
  n_basis: 8
  n_convo: 3
  n_harmo: 1
  even_gate: raw_swish
  even_act: raw_swish
  odd_act: tanh
  odd_gate: tanh
  radial_mlp_activa: raw_swish
  radial_mlp_hidden: 16
  radial_mlp_layers: 2
  self_connection: true
training:
  batch_size: 1
  charge_weight: 1.0
  energy_weight: 1.0
  forces_weight: 1.0
  smagmo_weight: 1.0
  magmom_weight: 1.0
  learning_rate: 0.0005
  max_epoch: 3000
  patience: 1000
\end{lstlisting}
In some cases when we noticed that the model didn't reached convergence after 3000 epochs the training was restarted by adding `restart: True` inside the training part of the configuration file.
Still, every model required a number of epoch for their training of the order of 10$^3$-10$^4$, consistent with the one reported in the orginal paper~\cite{LEOPOLD}.

In Supplementary Table \ref{tab:errors} the achieved RMSE for key quantities are reported for the final model used in the dynamics.
The errors achieved are in the range expected for Surface systems, while the errors on the polaron site are in line with the ones obtained for disordered systems, consistent with the effect of the polaron of reducing the symmetry of the system~\cite{Antoni2025,Zeng2025}.
Also, the increase in the force and magnetization error, with respect to the total ones, is consistent with the increase of $\sim$3-4 times reported in the original LEOPOLD paper~\cite{LEOPOLD} for bulk \ch{TiO2} in the presence of a Flourine defect.
\begin{table}[h]
	\centering
	\begin{tabular}{cccccc}
		\toprule
		           & $E$/Atoms & $F$                           & $F^\text{Pol}$               & $m$  & $m^\text{Pol}$ \\
		           & [meV]     & \multicolumn{2}{c}{[meV/\AA]} & \multicolumn{2}{c}{[m$\mu$]}                         \\
		\hline
		Train      & 0.196     & 20.9                          & 72.8                         & 3.49 & 57.9           \\
		Validation & 0.197     & 24.3                          & 109.9                        & 3.67 & 62.7           \\
		Test       & 1.94      & 22.9                          & 138.8                        & 2.30 & 37.1           \\
		\bottomrule
	\end{tabular}
	\caption{Root mean squared errors of the models for energy $E$, forces $F$ and magnetization $m$ and root mean squared errors of the models at the polaronic site for forces $F_\mathrm{pol}$, and the magnetization $m_\mathrm{pol}$. It's important to notice that the Test set has not been augmented during the active learning loop and thus posses a much lower statistical relevance compared to Validation and Train.}
	\label{tab:errors}
\end{table}

\section{Machine learning performances}

Every machine learning calculation from training to molecular dynamics was performed on the nodes of the LEONARDO HPC cluster, using one custom Ampere A100 GPU 64GB HBM2e, NVLink 3.0 (200GB/s) and a single socket 32-core Intel Xeon Platinum 8358 CPU, 2.60GHz (Ice Lake).
Using the specifcs of the model showed in the previous section the code allowed to generate 3~ns of trajectory on a 539 atom supercell in 10 hours, corresponding to 4.5$\times$10$^4$~atom-step/s comparable with other state of the art potentials.

%%%%%%%%%%%%%%%%%%%%%%%%%%%%%%%%%%%%%%%%%%%%%%%%%%%%%%%%%%%%%%%%%%%%%
%% BIBLIOGRAPHY
%%%%%%%%%%%%%%%%%%%%%%%%%%%%%%%%%%%%%%%%%%%%%%%%%%%%%%%%%%%%%%%%%%%%%
\printbibliography